\documentclass[12pt]{article}

\usepackage{amsfonts}

\def\form#1{(\ref{#1})}

\def\l{\lambda}

\def\r{\rho}
\def\s{\sigma}
\def\g{\gamma}
\def\al{\alpha}
\def\b{\beta}

\def\L{{\mathcal L}}
\def\G{{\cal G}}
 \def\U{{\mathcal U}}
 
 \def\R{{\cal R}}
\def\F{{\cal F}}
\def\H{{\mathcal H}}

\def\E{{\cal E}}

\def\F{{\mathcal F}}

\def\div{{\mathop{\rm Div}\nolimits}}
\def\tot{{\mathop{\scriptscriptstyle\rm Tot}\nolimits}}

\newcommand{\be}{\begin{equation}}
\newcommand{\ee}{\end{equation}}
\newcommand{\ba}{\begin{eqnarray}}
\newcommand{\ea}{\end{eqnarray}}
\newcommand{\baa}{\be\left\{\begin{array}{l}}
\newcommand{\eaa}{\end{array}\right.\ee}

\title{Energy for $N$--Body Motion in Two Dimensional
Gravity}
\author{R. B. Mann\thanks{E--mail: mann@avatar.uwaterloo.ca},
G. Potvin \\
Dept.\ of Physics, Univ.\ of Waterloo, Waterloo, Ontario
Canada N2L 3G1 \\
 \and and M. Raiteri \thanks{E--mail:
raiteri@dm.unito.it}\\ Dept.\ of Mathematics, Univ.\ of
Torino, via C.Alberto 10, 10123 Torino, Italy\\ }

\date{}

\begin{document}
\maketitle
\begin{abstract}
A general definition of energy is given, via the N\"other theorem,  for
the $N$--body problem in $(1+1)$ dimensional gravity. Within a
first--order Lagrangian framework, the density of energy of a
solution  relative to a background is identified with the
superpotential of the theory. For specific applications we
reproduce the expected Hamiltonian for the motion of $N$ particles
in a curved spacetime. This Hamiltonian
agrees with that found through an ADM--like
prescription for the energy when the latter is applicable but it
also extends  to a wider class of solutions provided a suitable
background is chosen.
\end{abstract}
%%%%%%%%%%%%%%%%%%%%%%%%%%%%%%%%%%%%%%%%%%%%%%%%%%%%%%%%%%%%%%%%%%%%%%%%%%%%%%%

\section{Introduction}

The study of  $2$--dimensional gravity has received much attention in
recent years motivated from  string--inspired  theories and from the
necessity  to study gravitational effects   in a simple mathematical
framework. It is well--known, for example,  that the problem of
$N$--body motion interacting by means  of their mutual gravitational
forces has no exact solution  in General Relativity.  This is
basically due to the dissipation of energy  through
gravitational
radiation. The problem considerably simplified in
two
dimensions where gravitational radiation is absent  but
the main features
of General Relativity are maintained.
Recently, a number of exact solutions
for $N=2$ were found in
\cite{RT,GTlet,GT,GTCharged}. These solutions have
qualitatively
different features compared to their non-relativistic
counterparts,
and provide a rich and interesting laboratory for the
study
of relativistic gravitational effects.

The $N$--body problem may be
formulated by taking  the matter action
to be  that of $N$ point particles
minimally coupled to gravity.
Because in  $(1+1)$--dimensional gravity the
gravitational field has
no real dynamical variables, the Lagrangian of the
theory must
include  some dynamics  through an auxiliary scalar field,
referred
 to as the dilaton in string theories.
Depending on the way the
dilaton field enters into the
Lagrangian, different models  arise. While we
develop
this formalism for an arbitrary dilaton theory of gravity,
we shall
be mainly interested in
three particular kinds of dilatonic gravity
theories,  
classified according the value of the equations of motion
impart to the
Ricci scalar $R$. They are the  Jackiw--Teitelboim
theory
\cite{JToriginal} (JT theory) with $R=\Lambda$, the
$R=T$ theory \cite{RT}
where T denotes the trace of the matter
stress--energy, and a more general
class of theories \cite{GT}
(encompassing the previous two  and called GT
theories from now on)
with $R=\Lambda+T$. In each of these theories the
evolution of the
gravitational field is governed only by the  matter
stress--tensor
(and vice versa) so that they mimic the features of
General
Relativity.

A general framework  for deriving the
Hamiltonian  for
the  $N$--particle system was developed in
\cite{RT,GT}. It was based on
the
$(1+1)$ counterpart of the ADM formalism; see e.g. \cite{Kimura}.
A canonical reduction of the action was carried out by eliminating
the Hamiltonian constraints and by imposing coordinate conditions.
The reduced Hamiltonian was defined  as the spatial integral of the
second spatial derivatives of the dilaton field. By solving the
constraints, the dilaton was given in terms of the coordinates and momenta
of the particles so that the reduced Hamiltonian is consequently a
function only of the parameters of the particles.
Moreover,  the consistency of the canonical
reduction  was proved in \cite{RT} and it was there shown that
the reduced Hamiltonian gives rise to equations of motion  which are
equal to the  original geodesic  equations for the
particles.

Nevertheless, the
key point in the definition of
the reduced Hamiltonian is the choice of the
coordinate and boundary conditions. 
Roughly speaking, these choices allow
one to discard all the boundary terms during the canonical reduction
of the action functional. If a solution does not satisfy the
required
conditions the reduced Hamiltonian can not be used to define
the
energy of the system. As  noted in \cite{RT}, the
definition of energy
in this situation becomes quite problematic and
new surface terms have to
be added ``ad hoc''; see \cite{gege}.

Motivated by this
kind of problem we
seek a definition of energy for  the
$N$--body problem not constrained by
boundary and coordinate
conditions.
The definition we shall present  in
this paper is based on the
N\"other theorem and  the theory of the
superpotential for
relativistic field theories;
see
\cite{FerrarisCQ,Robutti,Trautman,Remarks,OurBY} and references
quoted
therein. The approach we shall describe is essentially based  on
the
Lagrangian  formulation of a field theory and conserved quantities
are
defined with respect  to infinitesimal Lagrangian generators of
symmetries
on spacetime, namely, with respect to spacetime vector
fields. For each one
of these symmetries a superpotential (and
consequently an associated
conserved quantity) may be found
regardless of the topology of spacetime
and of the solution
considered. Superpotentials play a fundamental role in
the definition
of conserved quantities  since they enclose  the energetic
content of
the theory. For this reason we think that the specialization
of
N\"other theorem to the
$N$--body  problem in $2$--dimensional gravity
is well--suited
to define successfully a  generally valid expression for
the energy of
the system.

 The starting point for the construction of the
superpotential is the
definition of a {\it covariant}  action functional
for the theory
which is  {\it first--order} in the dynamical fields
(see
\cite{OurBY,Cavalese,Katz,TAUBNUT} and references quoted therein).
It
is obtained by adding a pure divergence  term to the
$N$--body
Lagrangian. Covariance is achieved by means of the
introduction of a
background solution chosen as a reference point (or
zero level) for
conserved quantities. The action functional so
obtained  is suitable for
taking field variations where only
the dynamical fields (and not their
derivatives) are kept fixed on the
boundary. In this way we have   a
Lagrangian which furnishes
a definition, via N\"other theorem, of a
conserved quantity which can be
truly considered as the  energy of the
system \cite{OurBY}.
Moreover, by  inserting the background from the very
beginning into
the Lagrangian, we naturally obtain the energy of the
solution
{\it relative} to the background.

We remark that the fixing of a
background  is a
universally valid procedure  because it does not imply
any restrictive hypothesis on the solution: the background  is
fixed depending on the solution under examination and for this reason it
is suited  to handle solutions ofwidely varying asymptotic behaviour
on the same footing; see
\cite{Remarks,Katz,TAUBNUT,HH,Hawking,BTZ1}.
Another advantage is that the energy of a region of {\it finite}
spatial extent  can be calculated through a suitable choice of
the background.\vspace{.6truecm}

The outline of our paper is as
follows.
In Section \ref{Lagrangian Formulation and ADM Hamiltonian}
the
Lagrangian formulation,  the ADM canonical reduction of the
theory and
the reduced Hamiltonian  for the  $N$--body system are
briefly summarized.
In Section
\ref{First--order covariant Lagrangian} the  Lagrangian
of
first--order in the dynamical fields is introduced and analysed.
In
Section
\ref{Superpotential} we devote considerable attention to the
construction of the superpotential starting from the first--order
Lagrangian. The
superpotential then furnishes a definition of the energy of a
solution relative to a background  contained in a spacelike region, i.e. a
real line interval.

In the rest of the paper we apply the formalism to
explicit
solutions. As far as we know the solutions here analysed cover
all the known exact (i.e., non--perturbative) relativistic solutions in
this dimensionality. In Section \ref{Applications: $R=T$ theory} we consider
the $N=2$ solution  for the JT theory. Inside our framework this is the
easiest theory  because a Minkowski--like background can be fixed for
the metric. This is not the case for the GT solutions  treated
in Section \ref{Application: GT theory}. The background here cannot
be Ricci flat and,
in order to avoid divergence problems,  it has to
be matched  with the
solution at the boundary  of the region of
integration. We shall also
investigate how the problem can be
simplified  by defining the {\it
variation} of energy  along a family
of solutions.

Section
\ref{Application: JT theory} is then devoted to  the JT
theory. Here the
dynamical metric  does not depend on the particle
coordinates (they are
instead contained in the dilaton field) so that
the dynamical metric  and
the background metric  are the same. While
the ADM recipe does not work
here for lack of the necessary
hypotheses,  the expected Hamiltonian is
reproduced via the N\"other theorem.

In the last Section  concluding remarks
and perspectives are presented.
%%%%%%%%%%%%%%%%%%%%%%%%%%%%%%%%%%%%%%%%%%%%%%%%%%%%%%%%%%%%%%%%%
%%%

\section{Lagrangian Formulation and ADM
Hamiltonian}
\label{Lagrangian Formulation and ADM Hamiltonian}

In order
to describe the motion of $N$ particles in
$2$D gravity we consider the
action
functional on the spacetime $M$:
\be
A=\int_M(\L_0+\L_P)\,d^2
x\label{totalactionone}
\ee
with
\ba
\L_0&=& {\sqrt{g}\over {2
\kappa}}\left[\psi\,R+{1\over
2}\,g^{\mu\nu}\,H(\psi)\,\nabla_\mu\psi\nabla_\nu\psi
+F(\psi) \right]\label{Lzero}\\
\nonumber \\
\L_P&=&\sum_a\int d\tau_a\left\{-m_a\left(-g_{\mu\nu}(x)
{dz_a^\mu\over d\tau_a}{dz_a^\nu\over
d\tau_a}\right)^{1/2}\right\}\delta^2(x-z_a(\tau_a))
\label{L_P}
\ea
where $R$ is the Ricci scalar of the metric $g_{\mu\nu}$, $g$
denotes
the absolute value of the metric determinant,
$\psi$ is  a scalar
field (the dilaton) and $H(\psi)$, $F(\psi)$ are
arbitrary functions of the
dilaton field not containing the
derivatives of the dilaton itself. Here
$\tau_a$ is the proper time
of the $a$--th particle and $\kappa=8\pi$ (in
geometric units with
$G=c=1$).

The variation of \form{Lzero} is given
by\footnote{Here and
in the sequel we use the notation 
$f(j^k \phi)$ to denote that the function
$f$ depends on the field $\phi$ together with
its
derivatives up to order $k$.}:
\ba
\delta\L_0&=&
{\sqrt{g}\over {2
\kappa}}\left[G_{\mu\nu}(j^2
\psi, j^1 g)\,\delta g^{\mu\nu}+G(j^2
g,j^2
\psi)\, \delta
\psi\right]+\nonumber\\
&&\label{varLzero}\\
&&+d_\mu\F^\mu(j^1g,
j^1\psi,
\delta (j^1 g),\delta (
\psi))
\nonumber
\ea
where:
\ba
&&G_{\mu\nu}=g_{\mu\nu}\nabla^\s\nabla_\s\psi
-\nabla_\mu\nabla_\nu\psi+{H\over 2}\left(\nabla_\mu
\psi
\nabla_\nu\psi-{1\over 2}g_{\mu\nu}
\nabla^\s\psi\nabla_\s\psi\right)
-{1\over 2}g_{\mu\nu}F\nonumber\\
&&G=R-H\,\nabla^\s\nabla_\s\psi-{1\over2}H'
\nabla^\s
\psi\nabla_\s\psi+F'\label{eom}
\ea
$H'$ and $F'$  denoting
the derivatives
with respect to the functional argument.
{The explicit expression of the
divergence terms in \form{varLzero}
will be given in   Section
\ref{Superpotential} -- see equation
\form{26}.

The field equations derived
from the action
\form{totalactionone}
are
\ba
&&G=0\label{eomofgone}\\
&&G_{\mu\nu}=\kappa\,T_{\mu\nu}\label{eomofg}\\
&&{d\over d\tau_a}\left\{ g_{\mu\nu}(z_a){dz_a^\nu\over
d\tau_a}\right\}-{1\over 2} g_{\nu\l ,\mu} (z_a)
{dz_a^\nu\over d\tau_a}{dz_a^\l\over d\tau_a}=0
\label{geodesic}
\ea
where $G$ and
$G_{\mu\nu}$ have been defined in \form{eom}
and
\be
T_{\mu\nu}=\sum_a m_a
\int d\tau_a {1\over \sqrt{g}}
g_{\mu\s}g_{\nu\r}{dz_a^\s\over
d\tau_a}{dz_a^\r\over
d\tau_a}\delta(x-z_a(\tau_a))
\ee
Since $\nabla^\mu
\,G_{\mu\nu}=-1/2\, \nabla_\nu\, \psi \, G$,
equations \form{eomofgone} and
\form{eomofg} together guarantee  the
conservation law $\nabla_\mu\,
T^{\mu\nu}=0$ when the equations of
motion are satisfied.

We next
consider how the three particular theories of interest
we noted above
arise.
\begin{description}
\item[1--] If we set $H=0$ and
$F=-\psi\,\Lambda$ into the
Lagrangian $\L_0$ we recover the
Jackiw--Teitelboim theory 
(JT theory)\cite{JToriginal}
for the
gravitational field coupled to $N$ point masses. 
In that case, equation
\form{eomofgone}
reduces to:
\be
R=\Lambda
\ee
The Ricci scalar is a
constant and the other dynamical fields evolve
in this spacetime of
constant curvature.
\item[2--] for $H=1$ and $F=0$ we obtain the so--called
$R=T$ theory;
see \cite{RT}. From \form{eomofgone} and \form{eomofg} we
have
\be
R=\kappa\,T^\mu_\mu\label{R=T}
\ee
In this theory the matter
affects the evolution of spacetime through
the trace of the stress energy
$T_{\mu\nu}$.
\item[3--] if we set  $H=1$ and $F=\Lambda$ we obtain the
generalised
theory (GT theory)  described in \cite{GT}. In this case
combining
the trace of equation \form{eomofg} with \form{eomofgone} we
obtain:
\be
R=\Lambda +\kappa\, T^\mu_\mu\label{GT}
\ee
When all bodies are
massless the GT theory reduces to the JT theory
whereas when the
cosmological constant vanishes we recover the
$R=T$
theory.
\end{description}
Other choices of the functions $H$ and $F$ allow
one to recover the
dilaton gravitational theories  studied in
\cite{BHMann}. As yet
there are no exact solutions to the $N$-body problem
in any
of these other theories.
\vspace{1truecm}

We shall now  derive the
canonical form for the 
action \form{totalactionone}. For simplicity, we
assume
$H=$const, a  choice   compatible with both
the JT theory ($H=0$)
and  the $R=T$ and GT theories
($H=1$). 

Our goal will be  to obtain the
definition of the Hamiltonian
in the ADM formalism (see \cite{RT,Kimura})
and to compare it  with
the definition of energy  we shall obtain  via
N\"other theorem and
the theory of superpotentials.

Let us then
consider
the ADM splitting of the  metric:
\be
g=-N_0^2 \, dt^2+\g\,\left(dx +
{N_1\over \g}\, dt
\right)^2\label{ADMmetric}
\ee
{}From now on, we shall
use the symbols $(\,\dot{}\,)$ and $({}')$ to
denote the
derivatives
$\partial_t$ and $\partial_x$, respectively.
  The
action
\form{totalactionone} transforms to:
\ba
A&=&\int_M d^2
x\left\{\sum_a p_a\dot z_a\delta (x-z_a(t))+\pi\dot \g
+\Pi\dot \psi+N_0
R^0+ N_1 R^1\right\}\label{ADMaction}\\
&& +\hbox{boundary
terms}\nonumber
\ea
where $\pi$ and $\Pi$ are the momenta conjugated to
$\g$ and $\psi$,
respectively
\ba
&&\pi= {1\over
2\kappa\sqrt{\g}N_0}\left(-\dot \psi +{N_1\over
\g}\psi'
\right)\nonumber\\
&&\Pi={1\over  2\kappa N_0}\left\{  -{1\over
\sqrt{\g}}\left( \dot \g+ N_1
{\g'\over \g}-2N_1'\right)
+\sqrt{\g}H(-\dot
\psi + {N_1\over \g}\psi')
\right\}\nonumber
\ea
and $R^0$ and $R^1$ are
the Hamiltonian and the momentum
constraints,
respectively:
\ba
R^0&=&-\kappa\, H \sqrt{\g}\g\,\pi^2+2
\kappa\sqrt{\g} \,\pi\,\Pi+ {H\over
4\kappa\sqrt{\g}}(\psi')^2-{1\over
\kappa}\left({\psi'\over
\sqrt{\g}}\right)'\nonumber\\
&&+{\sqrt{\g}\over
2\kappa}
F-\sum_a\sqrt{{p_a^2\over
\g}+m_a^2}\,\delta(x-z_a(t))\label{H0constr}\\
R^1&=&{
\g'\over \g}\,\pi-{1\over \g}\,\Pi\,\psi'+
2\pi'+\sum_a{p_a\over\g}\delta
(x-z_a(t))\label{H1constr}
\ea
The technique developed in \cite{RT} for
defining the Hamiltonian of
the system was to consider  the ``total
generator''  of the action
\form{ADMaction}. This procedure allowed one to
identify the dynamic and
the gauge character of the variables and,
subsequently,  to impose
the coordinate conditions
\be
\g=1\qquad
\Pi=0\label{coordconditions}
\ee
Eliminating then the constraints
\form{H0constr} and \form{H1constr}
and adopting the choice
\form{coordconditions},  the action
\form{ADMaction} then simplifies to the
{\it reduced} form:
\be
A_R=\int_M d^2x \left\{ \sum_a p_a \dot z_a \delta
(x-z_a) -\H\right\}
\qquad \H=-{1\over
\kappa}\psi''\label{reducedaction}
\ee
which is very similar to the
situation in classical mechanics.
The reduced Hamiltonian for the system of
particles is then
identified with:
\be
H=\int dx \,\H=-{1\over \kappa}\int
dx \,\psi''\label{ADMham}
\ee
where $\psi''$ is understood to be a function
of $z_a$ and $p_a$ by
solving the constraints \form{H0constr} and
\form{H1constr}. Despite
the very simple form of the Hamiltonian, {\it
appropriate boundary
conditions} have to be imposed in order to guarantee
the vanishing
of  extra boundary terms in the action \form{reducedaction}.
These
conditions will be analysed case by case in the next sections and
we
shall see that they play a fundamental role in establishing
the
equivalence between the reduced  Hamiltonian \form{ADMham} and
the
N\"other definition of
energy.
%%%%%%%%%%%%%%%%%%%%%%%%%%%%%%%%%%%%%%%%%%%%%%

\section{First--Order Covariant Lagrangian}
\label{First--order covariant Lagrangian}

In order to
calculate the energy via N\"other theorem of a system of
$N$ particles
coupled to gravity we
have to slightly modify the action functional
\form{totalactionone}
by adding a boundary term. As it is well known, the
addiction of
boundary terms into the action does not affect the equations
of motion
provided the correct boundary conditions are taken into account
in
order to cancel the boundary terms which arise from the bulk action
in
the variation.
Hence, variations with  prescribed boundary conditions
select suitable
boundary terms to append to the action and vice versa
\cite{OurBY,York,BrownYork}. When dealing with conserved
quantities,
additional boundary terms  lead to different values for the
N\"other
currents and for the associated  N\"other charges and each one has
a
different physical meaning. We shall here consider a modified
action
functional leading to the definition of a conserved quantity
which
can be ``truly'' considered as the energy of the system. It
is
constructed by requiring that, in the variational principle,
the
dynamical fields are kept fixed on the boundary while the variation
of
their derivatives are free, i.e. they are not constrained to
vanish.

Let us consider again
the variation $\delta \L_0$ in
\form{varLzero}. It is is easy to verify 
(see \form{26} below) that the divergence terms
depend not only on $j^1g$, $j^1\psi$,
$\delta g^{\mu\nu}$, $\delta \psi$ but also on
$\delta u^\al_{\b\mu}$
where
\be
u^{\al}_{\b\mu}=\Gamma^\al_{\b\mu}-
\delta^\al_{(\b}\Gamma^\nu_{\mu)\nu}\label{u}
\ee
{}From \form{varLzero} we then see that the action
functional
$A^0_M=\int_M \L_0 d^2x$ is stationary, i.e. $\delta A^0_M
=0$,
if the Euler--Lagrange equations of $\L_0$, i.e. $G_{\mu\nu}=0$
and
$G=0$, are satisfied and if $g^{\mu\nu}$ and $\psi$
are kept fixed on
the boundary
$\partial M$ together with certain  derivatives
 of
$g^{\mu\nu}$ (namely $\left.\delta
u^\al_{\b\mu}\right\vert_{\partial
M}=0$).
In order to have  a variational principle where {\it only}
the
dilaton field and the metric are fixed on the boundary of
the region of
integration (while the metric derivatives are not
fixed), we have to build
an action functional of first order in the
dynamical fields. As already
explained, this may be done by adding a divergence term to the
Lagrangian
\form{Lzero}. The ultimate motivation resides in the necessity
of having a well--posed definition for the energy of the system.

Moreover,when dealing with the theory of conserved quantities,  it
is well known that {\i
t absolute} conserved quantities
(for example the absolute energy of a
solution) do not
have a precise  meaning in General Relativity since it is
preferrable
to consider  {\it relative} conserved quantities;
see
\cite{TAUBNUT,Remarks,HH,Hawking,BTZ1,OurBY,BrownYork}. By
{\it
relative} we mean that the conserved quantities of a dynamical
field
are calculated with respect to a background solution which
is chosen
as a zero level or reference point.

In order to have a first order covariant Lagrangian with 
the background initially included in the
action functional we consider
a background metric $\bar g$. By denoting with
$\bar \Gamma^\al_{\b\mu}$ the Levi--Civita connection
of the background we may define
the {\it first--order covariant} Lagrangian:
\ba
\L_1&=&\L_0+\L_{\div}\nonumber\\
&=&{\sqrt{g}\over {2 \kappa}}\left(\psi\,R+{1\over
2}\,g^{\mu\nu}\,H\,\nabla_\mu\psi\nabla_\nu\psi
+F\right)\nonumber\\
&&-{1\over {2
\kappa}}d_\al(\sqrt{g}\,\psi\, g^{\mu\nu}w^\al_{\mu\nu})\label{Lone}
\ea
where
\be
w^\al_{\b\mu}=Q^\al_{\b\mu}-\delta^\al_{(\b}Q^\nu_{\mu)\nu}
\qquad
Q^\al_{\b\mu}=\Gamma^\al_{\b\mu}-\bar\Gamma^\al_{\b\mu}
\ee

A first glance  at \form{Lone} shows that the background
$\bar g$  does not affect the equations of motion
since it is contained in the divergence term $\L_{\div}$.
Moreover \form{Lone} is manifestly covariant  since
$w^\al_{\b\mu}$ is a tensor. We also observe that $\L_1$
is first--order in the dynamical fields $g$ and $\psi$ (and
second--order in the background $\bar g$). Indeed, the
terms
$\sqrt{g}\psi\,R-d_\al(\sqrt{g}g^{\mu\nu}
\psi\,w^\al_{\b\nu})$ may be rewritten as\footnote{Henceforth
all quantities with bars refer to background objects so
that, for example, $\bar R_{\mu\nu}$ denotes the Ricci
tensor of the background metric $\bar g$.}
\[
\sqrt{g}\left\{\psi g^{\mu\nu} \bar R_{\mu\nu}
-g^{\mu\nu}\nabla_\al\psi\, w^\al_{\mu\nu} +g^{\mu\nu} \psi
(Q^\al_{\s\nu}Q^\s_{\mu\al}-Q^\al_{\al\s}Q^\s_{\mu\nu})
\right\}
\]
Being first--order, the action functional $\int_M \L_1 d^2 x$  is
extremized by the dynamical fields $g$ and $\psi$ which
satisfy the equations of motion \form{eom} provided that
$\left.\delta g\right\vert_{\partial M}=0$ and
$\left.\delta \psi\right\vert_{\partial M}=0$.
Notice that, until now, the background has been introduced  only to
provide covariance for the Lagrangian $\L_1$. It also seems
reasonable  to
require  that the  background  $\bar g$ is a solution of field
equations without particles, i.e. a vacuum solution. For this reason,
in  describing  the motion of $N$ particles in
$2$--dimensional gravity one could  replace the action
functional \form{totalactionone} with:
\be
A^\tot_M=\int_M(\L_0+\L_\div+\L_P-\bar \L_0)\,d^2
x\label{totalaction}
\ee
where $\bar\L_0$ is the Lagrangian obtained from
\form{Lzero} by
substituting  {\it all} the fields (the metric $g$ and also
the
dilaton $\psi$) with the corresponding background fields $(\bar
g,\bar
\psi)$. In this way, the field equations  for 
$(\bar g,\bar\psi)$
are the same as those in \form{eomofgone} and
\form{eomofg}
with
$T_{\mu\nu}=0$.

We remark that a similar technique was employed in
General Relativity in order to calculate the corrected relative
conserved quantities for a large class of solutions; see
e.g. \cite{Katz,Cavalese,BTZ1,TAUBNUT,OurBY}.

%%%%%%%%%%%%%%%%%%%%%%%%%%%%%%%
%%%%%%%%%%%%%%%%%
\section{Superpotential}
\label{Superpotential}

In this Section we shall briefly review the theory of
conserved quantities via N\"other theorem. The formulation
we shall give here is based on the
geometric formulation of a field theory and it
basically relies on covariance requirements for the
Lagrangian describing the physical model. The detailed and
rigorous  construction  of N\"other currents, superpotentials
and conserved quantities requires the geometric
formulation of a relativistic field theory in terms of  fiber bundles and
their jet prolongations. We refer the interested reader to
\cite{FerrarisCQ,Robutti,Trautman,Remarks} and references quoted
therein for  details. Here we specialize the formalism developed in
those papers in order to construct the superpotential associated with
the action \form{totalaction}.

 First of all let us consider the  Lagrangian $\L_0$ in
\form{totalaction}. It is a covariant Lagrangian density, i.e. it
is invariant under coordinate transformation. This means
that, for any vector field $\xi=\xi^\mu\,  \partial/\partial
x^\mu$ on the spacetime $M$, the following identity holds (see
\cite{FerrarisCQ}):
\be
d_\l(\xi^\l\,\L_0)=\G_{\mu\nu} \,\pounds_\xi (g^{\mu\nu})+
\R^{\mu\nu} \,\pounds_\xi (R_{\mu\nu})+\H\,\pounds_\xi(\psi)+
\H^\mu\,\pounds_\xi(\nabla_\mu \psi)\label{variation}
\ee
where:
\ba
&&\G_{\mu\nu}={\partial\L_0\over
\partial g^{\mu\nu}}={\sqrt{g}\over 2k}\psi(R_{\mu\nu}-1/2
g_{\mu\nu}R)=0\nonumber\\
&&\R^{\mu\nu}={\partial\L_0\over
\partial R_{\mu\nu}}={\sqrt{g}\over 2k}\psi
g^{\mu\nu}\nonumber\\
&&\H={\partial\L_0\over \partial
\psi}={\sqrt{g}\over 2k}
(R+{1\over
2}g^{\mu\nu}\,H'(\psi)\,\nabla_\mu\psi\nabla_\nu\psi
+F'(\psi) )     \nonumber\\  &&
\H^\mu={\partial\L_0\over \partial
(\nabla_\mu\psi)}={\sqrt{g}\over
2k}g^{\mu\nu}\,H(\psi)\,\nabla_\nu\psi\nonumber\\
\nonumber
\ea
and $\pounds_\xi$ denotes the Lie derivarive with respect to
the vector field $\xi$.

 Taking into account the relation
$\pounds_\xi R_{\mu\nu}=\nabla_\al (\pounds_\xi
u^\al_{\mu\nu})$ (where
$u^\al_{\mu\nu}$ has been defined in \form{u}), through a
covariant
integration by parts
the identity \form{variation} can be rewritten as
(compare with \form{varLzero})
\be
d_\s\left\{\F^\s-\xi^\s\,\L_0\right\}=
-{\sqrt{g}\over {2 k}}\left[G_{\mu\nu}(j^2
\psi, j^1 g)\,\pounds_\xi
g^{\mu\nu}+G(j^2
g,j^2 \psi)\, \pounds_\xi
\psi\right]\label{notherth}
\ee
where
\ba
&&\F^\s(j^1g,
j^1\psi,
\pounds_\xi (j^1 g),\pounds_\xi ( \psi))
={\sqrt{g}\over
2k}\left\{g^{\mu\nu}\,\psi \pounds_\xi
u^\s_{\mu\nu}
+\right.\nonumber\\
&&\phantom{\F^\s(j^1g,}\left.-\nabla_\al\psi(g
^{\al\s}g_{\mu\nu}-
\delta^\al_\mu\delta^\s_\nu)\pounds_\xi
g^{\mu\nu}+H(\psi)\,
g^{\s\nu}\,\nabla_\nu\psi\pounds_\xi\psi\right\}\label{26}
\ea
The N\"other current $\E(\L_0,\xi) $ associated
with the generator $\xi$ of
Lagrangian symmetries is defined
as 
\be
\E^\s(\L_0,\xi) =\F^\s-\xi^\s\,\L_0\label{current}
\ee
It is a $(n-1)$--differential form on the spacetime $M$,
i.e. a $1$--form.

{}From \form{notherth} it is clear
that,
whenever $g$ and $\psi$ are solution of the field equations
obtained
from the Lagrangian $\L_0$, namely:
$G_{\mu\nu}=0$ and $G=0$, the
N\"other current $\E^\s(\L_0,\xi)$ obeys
the continuity equation:
\be
d_\s
\E^\s(\L_0,\xi)=0
\ee
Moreover, since the mapping $\xi\mapsto
\pounds_\xi(\cdot)$ is a
linear partial differential operator, the N\"other
current
\form{current} can be expanded as a linear combination of
the
symmetrized covariant derivatives of $\xi$ up to second
order:
\be
\E^\al(\L_0,\xi)=T^{\al}_{\mu}\xi^{\mu}+
T^{\al\b}_{\mu}\,\nabla_\b\xi^{\mu}+
T^{\al\b\g}_{\mu}\,\nabla_{(\b}\nabla_{\g)}\xi^{\mu}
\label{linear}
\ee

where the {\it canonical tensors} $T$
are:
\ba
&&T^{\al}_{\mu}={\sqrt{g}\over
2k}\left\{
H\,g^{\al\nu}\nabla_\nu\psi
\,\nabla_\mu\psi+{3\over
2}\psi\,R^\al_\mu-\delta^\al_\mu\,\L_0\right\}\nonumber\\
&&T^{\al\b}_{\mu}={\sqrt{g}\over
k}h^{\l\al\b}_\mu\nabla_\l\psi
\label{definitionT}\\
&&T^{\al\b\g}_{\mu}=-{\sqrt
{g}\over k}\psi
h^{\al(\b\g)}_\mu\nonumber
\ea
with
\[
h^{\s\r\g}_\nu=g^{\s\r}\delta^\g_\nu-g^{\
g(\s}\delta^{\r)}_\nu
\]
Whenever  we have a linear combination of the
form
\form{linear} we can perform a covariant integration by
parts  to
obtain for the same quantity  an equivalent
linear expansion  whose
coefficients  are all symmetric
with respect to upper indices, while the
integrated terms
are all pushed into a formal divergence \cite{Robutti}.
Doing so,
we obtain
\be
\E^\al(\L_0,\xi)=\tilde
\E^\al(\L_0,\xi)+d_\b\,
\U^{\al\b}(\L_0,\xi)\label{Notherzero}
\ee
with
\be
\tilde \E^\al(\L_0,\xi)=\tilde \E^{\al}_{\mu}\xi^{\mu}+
\tilde
\E^{\al\b}_{\mu}\,\nabla_\b\xi^{\mu}+
\tilde
\E^{\al\b\g}_{\mu}\,\nabla_{(\b}\nabla_{\g)}\xi^{\mu}
\ee
\ba
&&\tilde\E^{\al}_{\r}=T^{\al}_{\r}+t^{\al}_{\r}
-\nabla_\b\left(
T^{[\al\b]}_\r+t^{[\al\b]}_\r\right)
\nonumber\\
&&\tilde\E^{\al\b}_{\mu}
=T^{(\al\b)}_\r+t^{(\al\b)}_\r\label{Etilde}\\
&&\tilde\E^{\al\b\g}_{\mu}=T^{(\al\b\g)}
_\r\nonumber\\
\nonumber\\
&&t^{\al}_{\r}=1/3
\,T^{[\s\b]\al}_\g\,R^\g_{\r\s\b}\nonumber\\
&&t^{\al\b}_\r=-4/3\,\nabla_\s
T^{[\al\s]\b}_\r\nonumber
\ea
and
\be
\U^{\al\b}=\left\{ T^{[\al\b]}_\r
-{2\over 3}\nabla_\s
T^{[\al\b]\s}_{\r}
\right\}\xi^\s+{4\over
3}T^{[\al\b]\s}_{\r}\nabla_\s\xi^\r
\ee
The
$1$--form $\tilde\E(\L_0,\xi)$ is called {\it the reduced current}
while
the
$0$--form $\U(\L_0,\xi)=\U^{\al\b}\epsilon_{\al\b}$ is called {\it
the
superpotential} associated with the Lagrangian $\L_0$ and relative
to
the vector field $\xi$. It is easy to demonstrate
from
\form{notherth},
\form{current} and \form{linear} (see
\cite{Robutti,Remarks}) that the
reduced current is always vanishing
on--shell (i.e. when it is
evaluated on the solutions of the field
equations). In the present
case, from
\form{definitionT} and
\form{Etilde}
we obtain:
\ba
&&\tilde\E^{\al}_{\r}={\sqrt{g}\over k}
\,G^{\al}_{\r}
\nonumber\\
&&\tilde\E^{\al\b}_{\mu}=0\label{tildeE}\\
&&\tilde\E
^{\al\b\g}_{\mu}=0\nonumber
\ea
and
\be
\U(\L_0,\xi)={\sqrt{g}\over
2k}\left\{2\,\nabla_\b\psi\,
g^{\b\al}\,\xi^\s-\psi\,
g^{\al\mu}\,\nabla_\mu
\xi^\s
\right\}\,\epsilon_{\al\s}\label{supeL0}
\ee
A similar expression
can be found  for the Lagrangian $\bar\L_0$ in
\form{totalaction} by
replacing all the quantities involved in
\form{tildeE} and
\form{supeL0}
with the corresponding barred ones.

The same algorithmic technique can be
applied to the
Lagrangian $\L_{\div}$ in \form{totalaction} in order to
compute the
relative N\"other current $\E(\L_{\div}, \xi)$, the reduced
current
$\tilde\E(\L_{\div},\xi)$ and the superpotential $\U(\L_{\div},
\xi)$.
Because $\L_{\div}$ is a pure divergence, it gives no contribution
to
the reduced current and we have:
\be
\E(\L_{\div}, \xi)=d\,
\U(\L_{\div}, \xi)
\ee
where
\be
\U(\L_\div,
\xi)={\sqrt{g}\over
2k}\,\psi\,g^{\mu\nu}\,w^\b_{\mu\nu}\,\xi^\al
\,\epsilon_{\al\b}\label{Notherdiv}
\ee
Since our goal is to construct the N\"other
current $\E(\L,\xi)$
associated with the total Lagrangian
\form{totalaction}
\be
\L=\L_0+\L_{\div} +\L_P-\bar \L_0
\ee
we still have
to consider
 the $N$--particle matter  Lagrangian $\L_P$. It
 gives a
contribution $-\sqrt{g}
\,T^{\al}_{\r}$ to the reduced current while it
gives no contribution
to the superpotential.
Hence,  for the total
Lagrangian  $\L$,  we finally have
\be
\E^\al(\L,\xi)=\tilde\E^\al(\L,\xi)+
d_\b
\U^{\al\b}(\L,\xi)\label{nothertotal}
\ee
where
\be
\tilde\E^\al(\L,\xi)={\sqrt{g}\over \kappa}
\left\{\,G^{\al}_{\r}-\kappa\,T^{\al}_{\r}
\right\}\xi^\r
-{\sqrt{\bar g}\over \kappa}
\,\bar
G^{\al}_{\r}\,\xi^\r
\label{reduced}
\ee
and
\ba
\U(\L,\xi)&=&\U(\L_0,\xi)+\U(\L
_\div,\xi)-\U(\bar\L_0,\xi)\label{Utot}\\
&=&{\sqrt{g}\over
2\kappa}\left\{2\,\nabla_\b\psi\,
g^{\b\al}\,\xi^\s-\psi\,
g^{\al\mu}\,\nabla_\mu
\xi^\s
\right\}\,\epsilon_{\al\s}\\
&+&{\sqrt{g}\over
2\kappa}\left\{\psi\,g^{\mu\nu}\,w^\s_{\mu\nu}\,\xi^\al
\right\}\,\epsilon_{\al\
s}\label{Utwo}\\
&-&{\sqrt{\bar g}\over
2\kappa}\left\{2\,\bar\nabla_\b\bar\psi\,
\bar g^{\b\al}\,\xi^\s-\bar\psi\,
\bar
g^{\al\mu}\,\bar\nabla_\mu
\xi^\s
\right\}\,\epsilon_{\al\s}\label{superbarg}
\ea
{}From \form{eomofg} it is clear that the reduced current
\form{reduced}
vanishes on--shell, meaning that  the N\"other
current
\form{nothertotal} is not only a closed $1$--form on spacetime but
it
is also exact on--shell and  this is true independent of the
topology of
the spacetime. Moreover, while the current $\E(\L,\xi)$
is conserved just
along solutions  the quantity
$\E(\L,\xi)-\tilde\E(\L,\xi)$ is conserved
along every field
configuration, including those which are not solutions
of the field equations (see \form{nothertotal}). 
We express this fact  by
saying that
$\E(\L,\xi)$ is {\it weakly conserved} while the
difference
$\E(\L,\xi)-\tilde\E(\L,\xi)$ is  {\it strongly
conserved}.
Notice that these resultshold true for every vector field $\xi$
on the spacetime $M$, not necessarily a Killing vector field.

Now let $D$
be a $1$--dimensional space--like region of the spacetime
$M$ and let the
boundary
$\partial D$ of $D$ be  formed by two points $P_1$ and $P_2$.
Let
$\xi$ be a time--like vector field  and let us denote by the
pair
$g(x),
\psi(x)$ (and $\bar g(x),
\bar \psi(x)$) a solution of the
field equations \form{eomofgone} and
\form{eomofg}. According to
\cite{FerrarisCQ,CADM} we
define the energy
$E^{\tot}_D(\L,\xi)$ of the
solution
 {\it relative} to the background, contained in the region
$D$ and  relative to the vector field $\xi$
as 
\be
E^{\tot}_D(\L,\xi)=\left.\U(\L,\xi)\right\vert_{P_2}-
\left.\U(\L,\xi)\right\vert_{P_1}
\label{ENERGY}
\ee
In other words,  the superpotential
\form{Utot} corresponds to the {\it density} of energy.
The  definition
\form{ENERGY} of energy relies entirely on the
covariant formulation of the
theory and can be considered  as
the {\it covariant}  counterpart of the
ADM formulation, see
\cite{CADM,Sinicco}. In this context, rather then
starting from a
preferred ADM foliation of the spacetime into
hypersurfaces, the
starting point  is an  arbitrary surface $D$ and  a
non--vanishing
vector field $\xi$, the flow of which defines the local
time.

Note that   the energy
\form{ENERGY} is already normalized in
such a way it is zero when
computed on the background.
We also stress that
the region $D$ in  \form{ENERGY} is not
required to extend out to ``spatial
infinity''. As we shall see below,
we can compute, via the N\"other
theorem,  also the ``quasilocal''
energy, i.e. the energy contained in a
region of finite spatial
extent.

We turn now to specific applications of
this formalism to the
$N$-body
problem.

%%%%%%%%%%%%%%%%%%%%%%%%%%%%%%%%%%%%%%%%%%%%%%%%%%%%%%%%%%%%%%%%%%%
\section{Applications: $R=T$ Theory}
\label{Applications: $R=T$ theory}

We shall here present the  exact solution to the problem of the
relativistic
motion of two point masses found in \cite{RT} and we
shall see that the
definition of energy  \form{ENERGY} coincides with
the ADM Hamiltonian
\form{ADMham}.

If $z_1$ and $z_2$ (with $z_2<z_1$) denote the
positions of
the two particles, the $1$--dimensional slices
$\{t=const\}$ of the ADM
foliation of spacetime can be divided  in
the three regions $x<z_2$,
$z_2<x<z_1$ and $z_1<x$. As in
\cite{RT} we call them the $(-)$ region, the
$(0)$ region and the
$(+)$ region, respectively.

First of all it is
important to stress that in the $R=T$ theory  it
is necessary to impose the
boundary conditions
\be
\psi=\pm 2\,\kappa\,\chi \qquad
\chi'=\pi\label{boundaryRT}
\ee
which must hold in the $(-)$ region as well
in the $(+)$ region.
That condition, together with the coordinate
conditon \form{coordconditions}, allows one
to pass from the expression \form{ADMaction} to the final
expression \form{reducedaction} where all the
boundary terms are
discarded.

The solution  in the $(+)$ and $(-)$ regions is
\ba
&&\g=1\qquad
\Pi=0\label{RTone}\\
&&\nonumber\\
&&N_0=A\phi^2=\left\{\begin{array}{ll}
              A\phi_+^2\quad&(+)\, \hbox{region}\\
\\
              A\phi_-^2&(-)\, \hbox{region}\\
              \end{array}
\right.\label{RTtwo}\\
&&\nonumber\\
&&N_{1(+)}=N_{0(+)}-1\qquad
N_{1(-)}=-N_{0(-)}+1\label{RTthree}\\
&&\nonumber\\
&&\psi=-4\ln\vert\phi\vert=
\left\{\begin{array}{ll}
-4\ln\vert\phi_+\vert\quad&(+)\, \hbox{region}\\
              \\
-4\ln\vert\phi_-\vert\quad&(-)\, \hbox{region}\\
              \end{array}
\right.
\label{RTfour}\\
&&\nonumber\\
&&\pi=-{N_1'\over \kappa\,N_0}\label{RTfive}
\ea
where $\phi_+$ and $\phi_-$ are rather
complicated functions of
$x,z_1(t),z_2(t), p_1(t),p_2(t)$ (see
ref. \cite{RT} for details). Notice that the boundary
condition
\form{boundaryRT} is satisfied if we choose the  plus sign in
the
$(+)$ region and the minus sign in the $(-)$ region, respectively.

The
reduced Hamiltonian \form{ADMham} becomes:
\be
H=-{1\over \kappa}\int
dx
\,\psi''=-{1\over
\kappa}\left[\psi'_{(+)}-\psi'_{(-)}\right]\label{hamilR=T}
\ee
A detailed analysis of the result \form{hamilR=T}
was carried out in ref. \cite{RT} where it 
was shown that the canonical equations of motion
\ba
&&\dot z_a={\partial H\over \partial p_a}\nonumber\\
&&\dot
p_a=-{\partial H\over \partial z_a}\nonumber
\ea
derived from \form{hamilR=T} give rise to the original geodesic
equations \form{geodesic}.  It was  also shown that for sufficiently
small values of the Hamiltonian the trajectories of the particles can
be considered  as a relativistic perturbation  of the Newtonian case.

Let us now compute the
energy of the two--particle system  through
the definitions \form{Utot} and
\form{ENERGY}.
First of all we have to select a suitable background adapted
to the
theory under examination, i.e., a zero level for the energy.
We
select as a background the vacuum
solution obtained from the equation of
motion \form{R=T} by setting
$T_{\mu\nu}=0$ (no--particles). Hence, a
natural
choice is
\be
\bar g=\eta \rightarrow d\bar{s}^2=-dt^2+ dx^2\qquad
\bar \psi=0\label{backflat}
\ee
Taking into account the ADM splitting of
the metric \form{ADMmetric},
the background \form{backflat} and setting
$\xi=\partial/\partial t$,
the superpotential \form{Utot}
becomes:
\be
\U=-{1\over \kappa \sqrt{\g}}N_0\,
\psi'+2\,N_1\,\pi\label{ADM_U}
\ee
Employing the boundary condition
\form{boundaryRT} and the
equations  \form{RTone} and \form{RTthree} it can
be rewritten as:
\be
\U_{(+)}=-{1\over \kappa}\psi_{(+)}'\qquad
\U_{(-)}=-{1\over \kappa}\psi_{(-)}'
\ee
in  the $(+)$ and in the $(-)$
regions, respectively.
Hence the expression \form{ENERGY} for the
energy
\be
E^{\tot}=\U_{(+)}-\U_{(-)}=
-{1\over
\kappa}\left[\psi'_{(+)}-\psi'_{(-)}\right]
\ee
perfectly agrees with the Hamiltonian
\form{hamilR=T}.

%%%%%%%%%%%%%%%%%%%%%%%%%%%%%%%%%%%%%%%%%%%%%%%%%%%%%%%%%
\section{Application: GT Theory}
\label{Application: GT theory}

We shall here
consider two classes of solutions for the GT theory
\form{GT},  a single
particle solution and
a two--particle solution. When computing the energy
via the N\"other
theorem  a careful analysis of the background has to
be done in both cases. We stress  that in GT theories  the
vacuum solution (i.e. the solution in absence of particles which is  a
good representative for the background) can not be identified with the
flat metric
$\bar
g=\eta$ together with $\bar
\psi=0$, as in the previous Section. Indeed, in
the GT theory  the
field equations
\form{eomofgone}, \form{eomofg}, in
absence of particles,  reduce to
\ba
&&R=\Lambda\label{R=L}\\
&&g^{\mu\nu}\nabla_\mu\nabla_\nu
\psi=\Lambda
\ea
Hence we have to  choose as a background  metric
a constant curvature solution satisfying \form{R=L}. Moreover, in order
to avoid divergence problems which are commonly encountered  in
the computation of conserved quantities in non--Ricci flat
spacetime (see, e.g. \cite{HH,Hawking,BTZ1,TAUBNUT,OurBY}), the
dynamical fields and the backgrounds  have to be matched on the
boundary of the region $I$ whose
energy content we want to calculate. If the region is
described by the expression $I=\{-x_{0}\le x\le x_{0},
t=\hbox{constant}\}$ we then require
$ g(t,\pm x_{0})=\bar g(t,\pm x_{0})$ and $
\psi(t,\pm
x_{0})=\bar\psi(t,\pm x_{0})$. The ADM reduction of the
superpotential
\form{Utot} then simplifies as follows:
\be
\U=-{N_0\over
k\sqrt{\g}}\left( \psi '-\bar \psi'\right) +2\,
N_1 \left(\pi-\bar
\pi\right)\label{super-match}
\ee
%%%%%%%%%%%%%%%%%%%%%%%%%%%%%%%%%%%%%%%%%%%%%%
%%%%%%%%%%%%%%%%%

\subsection{One--Particle Solution}
\label{One--Particle Solution}

We shall now describe  a solution
of the GT theory  with a point
particle of mass $M$ at the origin. In this
case the equations of motion \form{eomofgone} 
and \form{eomofg} reduce to
\ba
&&R-g^{\mu\nu}\, \nabla_\mu \nabla_\nu
\psi=0\label{GTone}\\
&&R=\Lambda -4\,\pi\,M \delta(x)\label{GTtwo}
\ea
If
we  look for a solution of the form:
\be
g=-\al(x) dt^2+{1\over \al(x)}
dx^2\label{solutiong}
\ee
equations \form{GTtwo} becomes:
\be
-{d^2\al\over
dx^2}=\Lambda -4\,\pi\,M \delta(x)
\ee
It is satisfied by
\be
\al(x)=1+4\pi
\,M\vert x\vert -{1\over2}\,\Lambda\, x^2\label{alpha}
\ee
Accordingly,
from \form{GTone} we obtain
that
\be
\psi(x)=-\ln\left(\al(x)\right)+
C\,t\label{solutionpsi}
\ee
is a
solution for every choice of the constant $C$.
We now want  to compute the
energy of the solution relative   to a
background. We can choose as a
background the vacuum solution $M=0$
or a solution with a different mass
$m$ as well:
\ba
&&\bar g=-\bar\al(x) dt^2+{1\over \bar\al(x)}
dx^2\\
&&\bar \psi=-\ln\left(\bar \al(x)\right)+
C\,t
\ea
where
\be
\bar
\al(x)=1+4\pi\,(M-m)\, x_0+4\pi \,m\vert x\vert
-{1\over2}\,\Lambda\,
x^2\qquad x_0=\hbox{const}>0
\ee
Notice that the background solution
$\{\bar g,\bar \psi\}$ satisfies
the same equations \form{GTone} and
\form{GTtwo} (with  mass $m$)
and it is matched with the solution $\{ g,
\psi\}$ on the boundary of
the real line interval
$I=\{-x_{0}\le x\le
x_{0}, t=\hbox{constant}\}$. Hence, from
\form{super-match} we easily
obtain
\be
\U(\pm x_0)=\pm {1\over 2}(M-m)
\ee
and from \form{ENERGY} we
deduce that the {\it relative} energy
$E^{\tot}_I$ contained in the
region
$I$, as expected,  is equal to
\be
E^{\tot}_I=\U(x_0)-\U(-
x_0)=M-m
\ee
We outline that this result holds  for all {\it finite}
real
intervals (and, of course, also asymptotically).

We also stress that
the definition \form{ADMham} of the ADM
Hamiltonian, instead, does not lead
to the correct value if applied
to this solution (when $x_0$ goes to
infinity it gives $H=0$). This is
basically due to the fact that the
solution under examination has
been obtained without imposing {\it a priori
} coordinate and
boundary conditions. Nevertheless, with the coordinate
transformation:
\baa
\tau=t+{1\over AB}\ln (1+A \vert x\vert)-{1\over
2AB}\ln (1+2A
\vert x\vert -{1\over2}\,\Lambda\, x^2)\\
\\
\vert y\vert=
{1\over A}\ln (1+A \vert x\vert)\qquad \qquad A={k
M\over 4},\quad
B=\sqrt{1+(8\Lambda/ k^2M^2)}
\eaa
the solution \form{solutiong},
\form{alpha}, \form{solutionpsi}
transforms into  the solution found in
\cite{GT}, Appendix B. In
the new  coordinate system  the coordinate
conditions
\form{coordconditions} are satisfied together with the
suitable
boundary conditions (see
\form{boundaryGT} below). Hence the
definition \form{ADMham} may
be
applied.

%%%%%%%%%%%%%%%%%%%%%%%%%%%%%%%%%%%%%%%%%%%%%%%%%%%
\subsection{Two--Particle Solution}

The exact solution for the  two--particle GT theory
was found in
\cite{GT} where the motion of the particles was  deeply
analysed
starting from the definition \form{ADMham} of the Hamiltonian.
In
this case the relevant formula \form{ADMham} was obtained by
imposing
boundary conditions which are the generalization to the GT
theory  of the
conditions \form{boundaryRT}. In the present
case they
read:
\be
\psi^2-4\,\kappa\,\chi^2+2\,\Lambda\, x^2=C_{\pm}\,x
\qquad
\chi'=\pi\label{boundaryGT}
\ee
for the $(+)$ and $(-)$ regions. Here
$C_{\pm}$ are constants.
We refer the reader to \cite{GT,GTCharged} for the
details. The solution in the $(+)$ and $(-)$ regions is given
by
\ba
&&\g=1\qquad
\Pi=0\label{RGone}\\
&&\nonumber\\
&&N_0=A\phi^2=\left\{
\begin{array}{ll}
         A\phi_+^2\quad&\phi_+(x)=B\exp{(K_+ x/2)}\\
               \\
         A\phi_-^2&\phi_-(x)=C\exp{(-K_- x/2)}\\
              \end{array}
\right.\label{RGtwo}\\
&&\nonumber\\
&&N_{1(+)}={Y_+\over
K_+}\,(N_{0(+)}-1)\qquad
N_{1(-)}=-{Y_-\over
K_-}(N_{0(-)}-1)\label{RGthree}\\
&&\nonumber\\
&&\psi=-4\ln\vert\phi\vert=\left\{
              \begin{array}{l}
-4\ln\vert\phi_+\vert\quad\\
              \\
-4\ln\vert\phi_-\vert\quad\\
              \end{array}
\right.\label{RGfour}
\ea
where
\ba
&&Y_{\pm}=\kappa\left[  X\pm
1/4(p_1+p_2)\right]\nonumber\\
&&\nonumber\\
&&K_{\pm}=\sqrt{Y^2_{\pm}-\Lambda/2}
\label{RGfive}\\
&&\nonumber\\
&&\pi_{\pm}=-X\mp1/4(p_1+p_2)\nonumber
\ea
and $A, B, C$ and $X$ do not depend on the $x$ coordinate but
implicitly depend
on the time through the positions $z_a$and the
momenta $p_a$ of the particles.
Hence the ADM Hamiltonian \form{ADMham} becomes
\be
H(z_a,
p_a)=-{1\over \kappa}\int_{-x_{0}}^{x_{0}} dx
\,\psi''={2(K_++K_-)\over
\kappa}\label{energyGT}
\ee
where  $-x_0< z_2<z_1< x_0$.

We shall see
that the same expression can be obtained as well  as a
N\"other charge
starting from the definition \form{Utot}. As in the
$R=T$ theory of Section
\ref{Applications: $R=T$
theory} the link between the ADM Hamiltonian and
the N\"other energy
is established via the boundary conditions
\form{boundaryGT}. We stress
again that only if  these conditions are
satisfied the expression
\form{ADMham} can be considered  as the true
Hamiltonian of the
system (in fact,  the one particle solution of section
5.1 is an
example where \form{ADMham} can not be applied) . Instead,
the
formula based on the N\"other approach  holds in any case, provided
we
choose the suitable background. When dealing with the
explicit
solution
\form{RGone}--\form{RGfour}
it is very cumbersome to find
a suitable background matching
correctly  the solution on the boundary.
Hence we prefer to
consider the infinitesimal version of equation
\form{super-match}.
We  consider a whole family  of
solutions
\form{RGone}--\form{RGfour} and we take the variation
$\left.
\delta\U\right\vert_{x_0}$ of the density of energy $\U$  at
the
boundary
$\{-x_0,x_0\}$
along this family (see \cite{CADM,Remarks}).
Moreover,  we demand
that every solution of the family has the same
boundary value, i.e.
$\left .\delta g\right\vert_{x_0}=0$ and 
$\left
.\delta\psi\right\vert_{x_0}=0$. Hence, from \form{super-match} we
obtain
\be
\delta\U=-{N_0\over \kappa\sqrt{\g}}\,\delta  \psi '
 +2\, N_1
\,\delta\pi\label{super-variation}
\ee
and from the
expressions
\form{RGfour} and \form{RGfive} we have
\baa
\delta
\psi_{\pm}'=\mp
2(Y_{\pm}/ K_{\pm})\,\delta Y_{\pm}\\
\\
\delta
\pi_{\pm}=-(1/ \kappa)\,\delta Y_{\pm}
\eaa

Then
$\delta\U_{\pm}=\pm\,2/ \kappa\,\,\delta K_{\pm}$ and the
variation of
the energy
becomes:
\be
\delta
E^{\tot}=\delta\U_{+}-\delta\U_{-}={2\delta(K_++K_-)\over
\kappa}
\ee
The latter expression obviously leads to the
result
\be
E^{\tot}={2(K_++K_-)\over \kappa}+
\hbox{const}\label{EnergyGT}
\ee
in agreement  with \form{energyGT}. The
constant of integration  can be fixed arbitrarily inside the family
of solutions depending on the choice for the zero of energy.
\vspace{1 cm}

We remark that the same result \form{EnergyGT}
may be achieved  if
the GT theory is extended in order to include charged
bodies, see
\cite{GTCharged}.
The action integral for gravitational and
electric fields coupled
with  $N$ charged point masses leads to the same
expression
\form{Utot} for the superpotential. [This does not mean that
the
superpotential and consequently the energy are not affected  by
the
electric field.  On the contrary,  the superpotential is evaluated on
a
solution and the latter depends on the electric field through  the
electric
stress energy  $T_{\mu\nu}^{el}$ appearing in the equations
of motion.]
Moreover, since the solution \cite{GTCharged} has the
same  structure
as
\form{RGone}--\form{RGfive}, the same calculation of this section may
be
repeated step by step yielding the same result
\form{EnergyGT}.

%%%%%%%%%%%%%%%%%%%%%%%%%%%%%%%%%%%%%%%%%%%%%%%%%%%%%%%%%%%%%%
%%%%%%%
\section{Application: JT Theory}
\label{Application: JT theory}

We
shall here describe a two--particle solution for the
Jackiw--Teitelboim
theory. It is obtained, we remind, by setting $H=0$
and
$F=-\psi\,\Lambda$
in
\form{Lzero} so that  the equations of
motion
\form{eomofgone}--\form{geodesic} read as
follows:
\ba
&&R-\Lambda=0\nonumber\\
&&g_{\mu\nu}\nabla^\s\nabla_\s\psi-\nabla_
\mu
\nabla_\nu\psi+
{1\over2}g_{\mu\nu}\psi\,\Lambda=\kappa\,T_{\mu\nu}\label{JT
equations}\\
&&{d\over d\tau_a}\left\{
g_{\mu\nu}(z_a){dz_a^\nu\over
d\tau_a}  \right\}-{1\over2} g_{\nu\l ,\mu}
(z_a)
{dz_a^\nu\over d\tau_a}{dz_a^\l\over d\tau_a}=0\nonumber
\ea
The
solution we shall describe is obtained by considering
$\Lambda=-2/l^2$ (with $l$ a real constant).

In order to find the solution, the coordinate
conditions \form{coordconditions} were imposed together with the choice
$N_1=0$. Hence  \form{ADMmetric} simplifies  to
$g=-N_0^2 dt^2+dx^2$ and
the field equations become
\ba
&&\dot \pi+ N_0 \left[{\psi \over 2 \kappa
l^2}-{p_1^2\delta(x-z_1(t))\over
2\sqrt{p_1^2+m_1^2}}-{p_2^2\delta(x-z_2(t))\over
2\sqrt{p_2^2+m_2^2}}\right]+ {N_0'\psi'\over 2\kappa}=0\\
&&\psi''-{\psi\over l^2} +\kappa\left(\sqrt{p_1^2+m_1^2}\delta(x\!-\!z_1)+
\sqrt{p_2^2+m_2^2}\delta(x\!-\!z_2)
\right)=0\label{equationforpsi}\\
&&
2\pi' +p_1\delta(x-z_1)+p_2\delta(x-z_2)=0\\
&&N_0''-{1\over
l^2}N_0=0\label{equationforN0}\\
&&\dot \psi +2\kappa N_0\pi=0\\
&&\dot p_1+{\partial N_0\over \partial
z_1}\sqrt{p_1^2+m_1^2}=0\label{part-one}\\
&&\dot p_2+{\partial
N_0\over
\partial z_2}\sqrt{p_2^2+m_2^2}=0\label{part-two}\\
&&\dot
z_1-N_0{p_1\over\sqrt{p_1^2+m_1^2} }=0\label{part-three}\\
&&\dot
z_2-N_0{p_2\over\sqrt{p_2^2+m_2^2} }=0\label{part-four}
\ea
with $\pi=-\dot \psi/ (2\kappa N_0)$ and $\partial N_0/ \partial
z_a=\left.\partial N_0(x)/ \partial x\right\vert_{x=z_a}$.
The solution $(g,\psi)$ is then given by:
\ba
\g&=&1\qquad N_1=0 \qquad N_0=\cosh \left({x\over
l}\right)\label{g-JT}\\
\psi&=&A(t) \left[\sqrt{p_1^2+m_1^2}\cosh \left({x-z_1\over l}\right)+
\sqrt{p_2^2+m_2^2}\cosh \left({x-z_2\over l}\right)\right]\nonumber\\
&&+B\left[\sinh \left({x-z_1\over l}\right)+\sinh \left({x-z_2\over
l}\right)
\right]\label{psi-JT}\\
&&-{\kappa \,l\over 2}\left[\sqrt{p_1^2+m_1^2}\sinh
\left({\vert x-z_1\vert\over l}\right)
+\sqrt{p_2^2+m_2^2}\sinh \left({\vert x-z_2\vert\over
l}\right)\nonumber
\right]
\ea
with $A(t)=a_1\cos \left(t/l\right)+a_2\sin \left(t/l\right)$ and
$B$, $a_1$ and $a_2$ constants. The reduced  Hamiltonian \form{ADMham}
now  becomes
\be
H=-{1\over \kappa}\int^\infty_\infty\,\psi''=0\label{H=0}
\ee
and it  clearly gives rise to an unphysical result. Indeed from
equations
\form{part-one}--\form{part-four}  the Hamiltonian
is expected to  be of the
form:
\be
H(z_1,z_2,
p_1,p_2)=N_0(z_1)\,\sqrt{p_1^2+m_1^2}+N_0(z_2)\,\sqrt{p_2^2
+m_2^2}
\label{expectedH}\ee
Without entering into the details we only
remark that  the wrong
result
\form{H=0} is basically due to the ``bad''
asymptotic behaviour of the
solution. Instead, as already observed, the
definition of
energy as a N\"other charge  is not constrained by
asymptotic
conditions and then it is  a good candidate for defining
the
Hamiltonian of the system. In order to apply the
definitions
\form{Utot} and
\form{ENERGY}, let us then fix the  background
solution $(\bar g,
\bar\psi)$ by setting
$z_a=0$ and
$p_a=0$ into the
solution
\form{g-JT}, \form{psi-JT}. We observe that the metric $g$ does
not
depend on the particles coordinates so that  $\bar g=g$. Hence
the
second term in \form{Utwo} is equal to zero (because
$w^\al_{\b\mu}=0$)
and the superpotential
$\U$ reads as follows:
\be
\U={1\over
\kappa}\left[\psi N_0'-\psi' N_0\right]-{1\over
\kappa}\left[\bar \psi
N_0'-\bar\psi' N_0\right]\label{superJT}
\ee
Denoting by $I$ the region
$I=\{-x_{0}\le x\le x_{0},
t=\hbox{constant}\}$
the energy $E^{\tot}_I$
contained in the region $I$
becomes:
\ba
E^{\tot}_I&=&\int_{-x_0}^{x_0}{d\U\over dx}dx\nonumber
\\
&=&{1\over \kappa}\int_{-x_0}^{x_0}\left(\psi N_0''-\psi''N_0
\right)dx-
{1\over \kappa}\int_{-x_0}^{x_0}\left(\bar\psi
N_0''-\bar\psi''N_0
\right)dx\nonumber\\
&=&\int_{-x_0}^{x_0}
N_0\left(\sqrt{p_1^2+m_1^2}\delta(x-z_1)
+\sqrt{p_2^2+m_2^2}\delta(x-z_2)
\right
)dx\label{99}
\ea
where in the last equality we have
taken  into account
the equation of motions
\form{equationforpsi}
and
\form{equationforN0}.
Then, if the region $I$ is ``large'' enough, i.e.
$-x^0<z_2<z_1<x^0$
the energy $E^{\tot}_I$ is
given by
\be
E^{\tot}_I=N_0(z_1)\,\sqrt{p_1^2+m_1^2}+N_0(z_2)\,\sqrt{p_2^2+m_2^2}
\ee
in agreemet with the expected value \form{expectedH}.

More
generally, note that the energy \form{99} is instead
given
by:
\baa
E=N_0(z_2)\,\sqrt{p_2^2+m_2^2} \qquad
\hbox{if}\quad
-x^0<z_2<x^0<z_1\\
\\
E=N_0(z_1)\,\sqrt{p_1^2+m_1^2} \qquad
\hbox{if}\quad
z_2<-x^0<z_1<x^0\\
\\
E=0 \qquad
\qquad\qquad\qquad\quad\hbox{if}\quad  z_2<-x^0<x^0<z_1
\eaa

Finally, we
point out that the above solution is easily extendable to 
$N$ particles,
since the spacetime is of constant curvature, and the metric
is given by
(\ref{g-JT}).  The solution for the $a$-th particle
is
\be
N_0(z_a)\sqrt{p^2_a+m^2_a}=E_a
\ee
where $E_a$ is a constant
and
\be\label{Znsoln}
z_a(\tau_a) = l
\cosh^{-1}\left[E_a\sqrt{\frac{2}{m_a^2+E_a^2+(m_a^2-E_a^2)\sin(\tau_a/l)}}
\right]
\ee
The total energy is
\be
E = \sum_{a=1}^N E_a
\ee
provided the region of interest surrounds all of the
particles.

%%%%%%%%%%%%%%%%%%%%%%%%%%%%%%%%%%%%%%%%%%%%%%%%%%%%%%%%%%%%%%%%%%%%
%%%%%%%%%%%%
%
\section{Conclusions and Perspectives}
\label{Conclusions and Perspectives}

We have analysed and compared two definitions of energy for the
$N$--body problem in $2$--dimensional gravity. They look   very
different  from a theoretical as well from a practical point of view.
Although the formula \form{ADMham} is very simple and easy to deal with
applications, we have shown in two examples (see Section
\ref{One--Particle Solution}  and Section \ref{Application: JT
theory}) that  it is
not generally valid. We have pointed out how this is related to
coordinates choices and boundary conditions. Instead, the definition
\form{ENERGY} of energy via N\"other theorem leads, in all the
situations so far analysed, to the expected value for the
Hamiltonian  and gives rise  to the same results of the ADM
prescription \form{ADMham} when the latter is applicable. The
definition \form{ENERGY} seems to be more difficult in applications
mainly because its $(1+1)$  ADM  splitting is not always the same,
compare e.g. \form{ADM_U},  \form{super-match} and
\form{superJT}. This is basically  due to the fact that the
background has to be chosen  in a different way every time. In our
opinion, it is exactly  the presence of the background inside its
expression which renders the definition  \form{ENERGY}  more general,
at least at a theoretical level. It is suitable for dealing   with
theories admitting  Minkowski--like  backgrounds (such as $R=T$ theories)
as well as more general contexts (GT and JT theories).

Divergence problems  in the energy are here cured  not imposing {\it
a priori} boundary
conditions (as it is commonly done in  the ADM canonical reduction)
and hence  fixing  the asymptotic behaviour  of the solution. Instead,
they are avoided, {\it a posteriori}, through the choice  of the
backgroud which suitably  matches the  solution under examination.

We expect that the definition  of energy
based on the superpotential  could be
applied also in  extensions of the formalism with additional
matter fields and to
$N$--body problem in $(2+1)$ dimensions, see e.g.
\cite{Matschull,Bellini}.

Finally, we stress that $2$ dimensional gravity, far from being a
pure mathematical toy model,  has deep relationships  with
dimensionally reduced four--dimensional spherically symmetric gravity
 and $2$D string--theoretical black hole gravity,
see \cite{Lau}, \cite{BHMann} and references quoted therein.
$2$D gravity
provides an important arena  for examining the notion  of
gravitational energy, and we anticipate it has more to teach
us about this
and other interesting
subjects.

%%%%%%%%%%%%%%%%%%%%%%%%%%%%%%%%%%%%%%%%%%%%%%%%%%%%%%%%%%%%%%
\section{Acknowledgments}
We are grateful to  L. Fatibene, M. Ferraris and M. Francaviglia   of
the University of Torino  for useful discussion on the subject,
and to T. Ohta of Miyagi University for helpful correspondence.
This work was supported by
INdAM--GNFM, by the  University of Torino (Italy), by the
University of Waterloo (Canada) and by the
Natural Sciences and Engineering Research Council of Canada.

%%%%%%%%%%%%%%%%%%%%%%%%%%%%%%%%%%%%%%%%%%%%%%%%%%%%%%%%%%%%%%%%%%%%%%%
\end{document}